\documentclass[12pt,preprint]{aastex}

\shorttitle{XMM-Newton phase-resolved spectroscopy of the Vela
pulsar} \shortauthors{Manzali, De Luca and Caraveo}

\begin{document}


\title{
Phase resolved spectroscopy of the Vela pulsar with XMM-Newton
\footnote{Based on observations with XMM-Newton, an ESA science
mission with instruments and contributions directly funded by ESA
member states and the USA (NASA).}}


\author{A. Manzali\altaffilmark{1}, A. De Luca, P. A. Caraveo}
\affil{Istituto di Astrofisica Spaziale e Fisica Cosmica, INAF \\
v. Bassini 15, I-20133 Milano, Italy}
\email{armando@iasf-milano.inaf.it,deluca@iasf-milano.inaf.it,pat@iasf-milano.inaf.it}

\altaffiltext{1}{Universit\`a degli Studi di Pavia, Dipartimento di Fisica
Nucleare e Teorica, Via Bassi 6, 27100 Pavia, Italy}



\begin{abstract}
The $\sim10^4$ y old \object{Vela Pulsar} represents 
the bridge between the young
Crab-like and the middle-aged 
rotation powered pulsars. Its
multiwavelength behaviour is due to the superposition of different
spectral components. We take advantage of the unprecedented harvest of
photons collected by
\emph{XMM-Newton} to assess the \object{Vela Pulsar} spectral shape and to
study the pulsar spectrum as a function of its
rotational phase. In order to fully exploit the data collected by
\emph{XMM-Newton} on the Vela pulsar,
we had first to discriminate the pulsar emission from that of the
bright surrounding nebula. To this aim, we used the
\emph{Chandra}/HRC surface brightness map of the nebula, coupled
with the most accurate calibration of the EPIC point spread
function. This procedure made it possible to assess the pulsar
spectral shape, disentangling its thermal component from the non
thermal one. As for the middle-aged pulsars Geminga, PSR B0656+14 
and PSR B1055-52 (the ``Three Musketeers''), the phase-integrated
spectrum of Vela is well described by a three-component model,
consisting of two blackbodies ($T_\mathrm{bb}=1.06\pm0.03\times
10^6\ \mathrm{K}$, $R_\mathrm{bb}=5.1^{+0.4}_{-0.3}\ \mathrm{km}$,
$T_\mathrm{BB}=2.16^{+0.06}_{-0.07}\times 10^{6}\ \mathrm{K}$,
$R_\mathrm{BB}=0.73^{+0.09}_{-0.07}\ \mathrm{km}$) plus a
power-law ($\gamma=2.2^{+0.4}_{-0.3}$). The relative
contributions of the three components are seen to vary as a
function of the pulsar rotational phase. The two blackbodies have
a shallow $\sim7-9\%$ modulation.
The cooler blackbody, possibly related to the bulk of the neutron
star surface, has a complex modulation, with two peaks per
period, separated by $\sim0.35$ in phase, the radio pulse
occurring exactly in between. The hotter
   blackbody, possibly originating from a hot polar region,
has a nearly sinusoidal modulation, with a single, broad maximum
aligned with the second peak of the cooler blackbody, trailing the radio pulse
by $\sim0.15$ in phase.
  The non thermal component, magnetospheric in origin, is present
only during 20\% of the pulsar phase and appears to be
opposite to the radio pulse. XMM-Newton phase-resolved
spectroscopy unveils the link between the thermally emitting
surface of the neutron star and its charge-filled magnetosphere,
probing emission geometry as a function of the pulsar rotation.
This is a fundamental piece of information for future
3-dimensional modeling of the pulsar magnetosphere.
\end{abstract}





\section{Introduction}
The \object{Vela Pulsar} is one of the best scrutinized neutron
stars. It was the first pulsating radio source observed in the
southern hemisphere \citep{1968Natur.220..340L} and the swing of
 the polarisation vector  during the
radio pulse provided evidence for a rotational origin of the
radio emission \citep{1969Natur.221..443R}. Moreover 
the position of the pulsar, close to the
centre of the \object{Vela Supernova Remnant} (hereafter SNR)
confirmed  the association of pulsars with rotating neutron stars
born from massive stars' collapse.

About ten years after the radio discovery, \cite{1977Natur.266..692W} detected a pulsating optical source of $V\sim23.6$: at least four peaks are present in the optical \citep{1998nspt.conf..363G} and UV \citep{2005ApJ...627..383R} light curves, as well as in the hard X-ray energy range \citep{2002ApJ...576..376H}. \emph{HST} observations \citep{2001ApJ...561..930C} of the optical counterpart have allowed the direct measurement of the pulsar distance ($294^{+76}_{-50}\ \mathrm{pc}$), later confirmed and refined through radio observations \citep[$287^{+19}_{-17}\ \mathrm{pc}$][]{2003ApJ...596.1137D}. \object{PSR\ B0833-45} is also a bright $\gamma$-ray source: pulsations were detected by \emph{SAS-2} \citep{1975ApJ...200L..79T}, \emph{COS-B} \citep{1977A&A....61..279B, 1980A&A....90..163K, 1988A&A...204..117G} and \emph{CGRO}. \cite{1994A&A...289..855K}, using EGRET, were able to perform spectroscopic analysis of different phase intervals of the double peaked $\gamma$-ray pulsar emission.

\emph{ROSAT} detection of soft X-ray pulsations
\citep{1993Natur.361..136O} was the last piece of the Vela
multiwavelength puzzle.  It turned out to be a difficult
observation since X-rays from the neutron star are embedded in
the bright Pulsar Wind Nebula (hereafter PWN), located near the
centre of the Vela SNR. A blackbody with temperature
$\sim1.5-1.6\times 10^{6}\ \mathrm{K}$ and a radius of $3-4\
\mathrm{km}$ could describe \emph{ROSAT} spectrum while the
nebular emission, clearly non-thermal, could be ascribed to
synchrotron emission originated from the interaction of the
high-energy particle pulsar wind with the interstellar medium.
\emph{Chandra} observations of the \object{Vela pulsar} provided
high resolution images of the X-ray nebula surrounding the
neutron star. The nebula turns out to be quite complex, with
spectacular arc and jet-like features \citep{2001ApJ...556..380H,
2003ApJ...591.1157P} reminiscent of those observed around the
Crab. The thermal nature of the pulsar spectrum inferred from
\emph{ROSAT} data was confirmed. While high resolution spectra
failed to show absorption features, a non-thermal harder tail was
found in ACIS-S spectra \citep{2001ApJ...552L.129P}. Analysing
\emph{XMM-Newton} observations \cite{2004AdSpR..33..503M} found
the same thermal radiation. However, in their preliminary
analysis, 
they studied only the pulsar phase-averaged thermal emission
below $\sim1$ keV.
Since thermal emission is a very distinctive
character of middle aged pulsars \citep[\object{Geminga},
\object{PSR\ B0656+14} and \object{PSR\ B1055-52},
][]{1997A&A...326..682B}, Vela's ($\tau\approx11\,400\
\mathrm{y}$ ; $\dot{E}\approx7\times 10^{36}\ \mathrm{erg}\
\mathrm{s}^{-1}$) spectral properties make it more similar to older
specimen than to younger ones. In this paper, we will provide
further evidence in favor of this classification, taking advantage
of \emph{XMM-Newton} statistics to perform phase-resolved
spectroscopy of the \object{Vela pulsar}.

\section{The Data}

We will use data collected by \emph{XMM-Newton} and
\emph{Chandra},  exploiting the characteristics of both
observatories. While \emph{XMM-Newton} observations provide an
unprecedented harvest of photons from the Vela pulsar and its
 PWN, the instrument point spread
function (hereafter PSF, 6.6\arcsec{} FWHM), however, does not
allow to disentangle the pulsar emission from the extended one.
Accounting for the bright PWN contribution is mandatory to unveil
the PSR emission properties. Thus, the high resolution image
obtained by the \emph{Chandra} observatory is crucial to have a
clear view of the Vela pulsar and its sorrounding nebular
emission.

\subsection{\emph{XMM-Newton} EPIC Data}

The X-ray Multi Mirror Mission observed \object{PSR\ B0833-45} on
2-3 December 2000. Owing to the source high flux, to reduce
photon pile-up pn and MOS2 were used in ``Small Window'' mode,
while MOS1 operated in ``Large Window''. In the \emph{XMM-Newton
Science Archive} there are two data sets (id 0111080101 and
0111080201), separated by about 1.5 h, that cover a time span of
(respectively) $41\,100\ \mathrm{s}$ and $61\,800\ \mathrm{s}$.
The raw data were first processed with standard pipeline tasks  of the XMM 
Science Analysis Software
(SASv6.5.0) and then the soft proton flares were removed using the
standard prescription of \cite{2004A&A...419..837D}. The two data
sets were merged using the SAS task \emph{merge}.
Table~\ref{table:time} summarise Good Time Intervals and Total
Exposure time for the three EPIC detectors.

\begin{figure}
\includegraphics[angle=0,width=10cm]{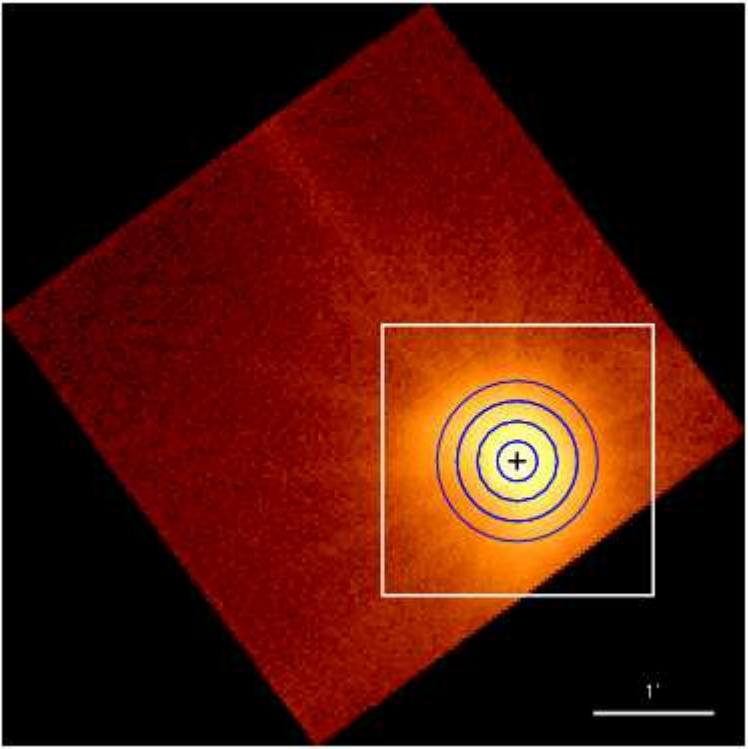}
\caption{EPIC/pn image of the \protect\object{Vela pulsar}. The cross
marks the radio position and the box marks the region covered by
\emph{Chandra} image of Fig.~\ref{fig:hrc}. The spectral
extraction regions (Sect.~\ref{par:spectra}) are also
marked.\label{fig:EPN}  }
\end{figure}

Fig.~\ref{fig:EPN} 
shows an image of the entire field of view, extracted using 
pn data: the compact nebula surrounding the \object{Vela pulsar},
with its peculiar shape, clearly stands out. 
The \emph{XMM-Newton} telescopes angular resolution, however, does
not allow to resolve the PSR from the PWN.
Indeed no spatial selection can disentangle the
point source from the diffuse emission.


\subsection{\emph{Chandra} HRC and ACIS Data}
\begin{figure}
\includegraphics[angle=0,width=10cm]{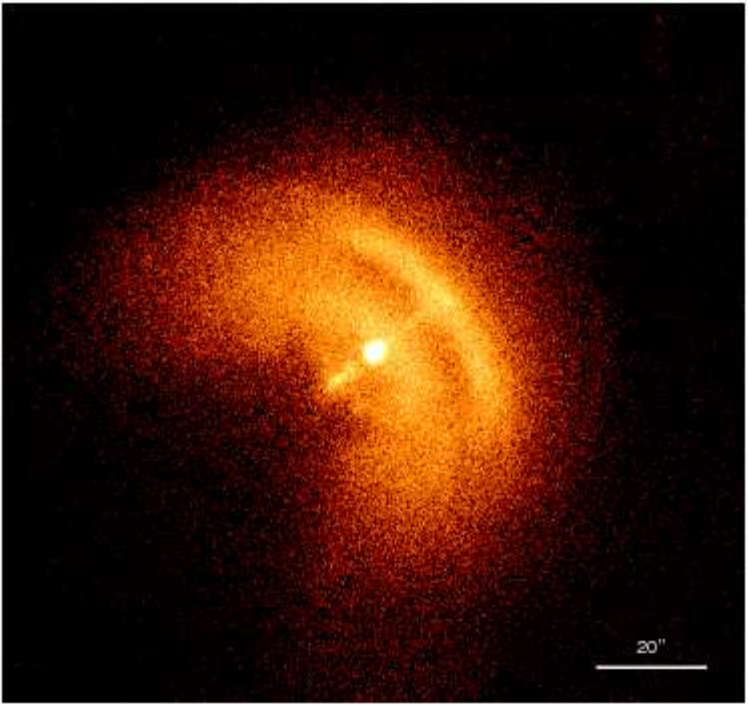}
\caption{\label{fig:hrc} \emph{Chandra}-HRC image of the \protect\object{Vela pulsar}.}
\end{figure}

To get a clear view of the spatial distribution of the bright
nebular emission, we analysed the two \emph{Chandra} HRC-Imaging
archive observations of the \object{Vela pulsar}, performed after
the January 2000 glitch. A third observation, taken on January
2002 was also available. All these observations were reprocessed
with CIAO\footnote{Chandra Interactive Analysis of Observations
(CIAO), \texttt{http://cxc.harvard.edu/ciao/}} v3.2, applying
aspect solution
\footnote{\texttt{http://cxc.harvard.edu/ciao/threads/arcsec\_correction/}}
and degap
correction\footnote{\texttt{http://cxc.harvard.edu/ciao/threads/hrci\_degap/}},
as well as reducing tap-ringing
distortion\footnote{\texttt{http://cxc.harvard.edu/ciao/threads/hrc\_ampsf/}}.
The resulting image (Fig.~\ref{fig:hrc}) shows the well known
structure of the Vela PWN: the two ``arcs'' and the
``jet/counter-jet'' feature protruding from the point source
\citep{2001ApJ...556..380H, 2003ApJ...591.1157P}.

\section{Spectral Analysis}
To perform the spectral analysis of the \emph{XMM-Newton} data on the Vela pulsar, we exploit both the spectral and temporal resolution of the EPIC/pn camera. To disentangle the PSR photons from the PWN ones, we take advantage of their different space distribution: while the PSR photons follow the instrument PSF, the nebular ones do not.

To estimate the nebular contribution in the EPIC/pn event file,
\begin{itemize}
\item{we extract spectra from concentric annular regions of increasing radii;}
\item{we determine the PSR contribution in each spectrum as a function of the PSF and of the Encircled Energy Fraction (EEF);}
\item{we determine the PWN contribution in each spectrum using a surface brightness map of the nebular emission derived from \emph{Chandra} data;}
\item{we fit all the spectra at once with a two component (PSR + PWN) model whose normalisation ratios are fixed following the two previous steps.}
\end{itemize}

In the following section we will describe in detail each step.

\subsection{EPIC/pn Phase-Averaged Spectra}
\label{par:spectra}

\begin{figure}
\includegraphics[angle=270,width=10cm]{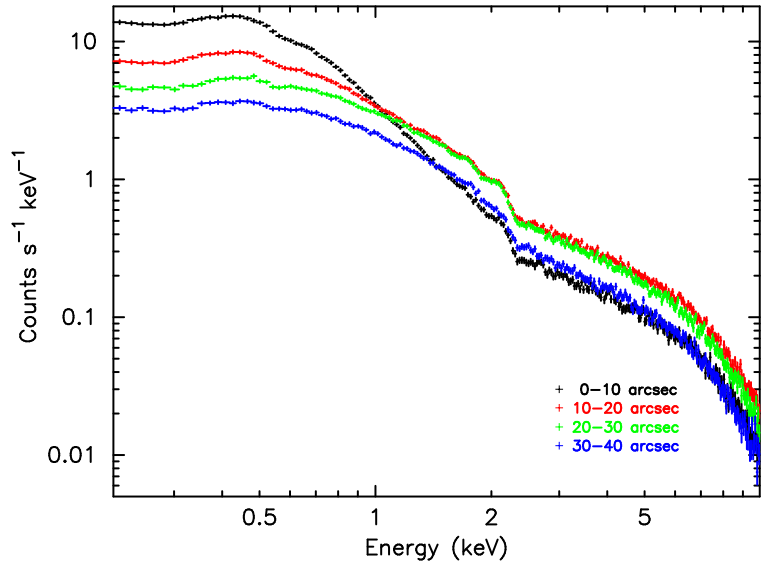}
\caption{\label{fig:spectra} Observed EPIC/pn spectra from the 4
different extraction regions.}
\end{figure}

We extract spectra from the merged EPIC/pn event  file, with FLAG
0 and PATTERN 0--4, from a  10\arcsec{} radius circle centered on
the pulsar position as well as from three annular
(10\arcsec--20\arcsec, 20\arcsec--30\arcsec,
30\arcsec--40\arcsec) regions (see Fig.~\ref{fig:EPN}). No
evidence for photon pile-up is found in the data.  All the
spectra are shown in Fig.~\ref{fig:spectra} where their composite
nature is clearly visible: while at lower energies the thermal
spectrum prevails, with the flux values in the 4 different
extraction regions proportional to EEF, at higher energies almost
all the photons detected are produced by the PWN and the flux
values are proportional to the spatial distribution of the nebula
surface brightness. Since background estimate will be provided by
the spectral fitting, we avoid background subtraction.  We chose the
energy range $0.2-10\ \mathrm{keV}$  and we rebinned the spectra
in order to have at least 30 counts per spectral bin and no more than 3 spectral 
bins per energy resolution interval. Owing to the extended nature of
the diffuse nebular emission we generated effective area files
with the prescription for extended sources, without modelling the
PSF distribution of the source counts.

\subsection{EEF From \emph{XMM-Newton} telescopes PSF} \label{par:psreef}
The \emph{XMM-Newton} telescopes radially averaged PSF profile at distance $r$ from the aimpoint is well described by the King's profile \citep{1998SPIE.3444..278G}:
\begin{equation}
PSF(r) = A \left[ 1 + \left( \frac{r}{r_{c}} \right)^2 \right] ^ {-\alpha},
\label{eq:king}
\end{equation}
where the core radius $r_{c}$ and the slope $\alpha$ vary as a function of the photon energy and off-axis angle and $A$ is a constant normalisation factor. The calibration files provide the values of $r_c$ and $\alpha$, for different photon energies and off-axis angles.

Since the \object{Vela pulsar} thermal spectrum peaks at about $0.5\ \mathrm{keV}$, and the source is always at the centre of the field-of-view, we used the parameters $r_c=1.59$ and $\alpha=5.504$, corresponding to an energy of $475\ \mathrm{eV}$ and an offset angle of 0.0 degrees.

The EEF, i.e. the fraction of energy collected within a certain radius $R$ from the source distribution centroid, is simply the integral of (\ref{eq:king}) times the radius over the radius, normalised at 5\arcmin:
\begin{equation}
EEF(R)= \frac{1 - \frac{1}{\left[ 1 + \left(\frac{R}{5.504'} \right) ^2 \right] ^{1.59-1}}}{1 - \frac{1}{ \left[ 1 + \left( \frac{5 '}{5.504'} \right) ^2 \right] ^{1.59-1}}}.
\label{eq:ee}
\end{equation}

Since the source is not strictly monochromatic, we calculated the fraction of energy encircled in the spectra extraction regions also for $250\ \mathrm{eV}$  and $800\ \mathrm{eV}$, in order to obtain a ``confidence interval''.

\subsection{Surface Brightness Map of the PWN From \emph{Chandra} Data}
\label{par:pwneef}
We used \emph{Chandra} data to produce a surface brightness  map
of the nebular emission. In the \emph{Chandra}/HRC images more
than $97\%$ of the pulsar counts falls within 3\arcsec{} from the
centroid of the distribution \citep[and references
therein]{2002PASP..114....1W}. Thus, outside this small region,
the images represent a good approximation of the brightness
spatial distribution in the energy range $0.1-10\ \mathrm{keV}$ .

Since we needed an estimate of the PWN flux  in the proximity of
the position of the pulsar, a two dimensional model of a
point-like source flux distribution was obtained with a
simulation performed with
ChaRT\footnote{\texttt{http://cxc.harvard.edu/chart/}} and
MARX\footnote{\texttt{http://space.mit.edu/CXC/MARX/}}. A 2D
fitting was then performed on the HRC images with the addition of
a spatially constant contribution from the PWN. Best fit count
rates found for the region between $2.64\arcsec$ and
$3.46\arcsec$, corresponding to $20-26\ \mathrm{pixel}$, are
reported in table~\ref{tab:PWN} for the three different HRC-I
exposures. Following the criterion by \cite{2001ApJ...556..380H},
the region inside $3.46\arcsec$, $26\ \mathrm{pixel}$, was excised
from all images and replaced with a poissonian distribution with a
mean value equal to the best fit count rate.

In order to take into account the angular resolution of  the
\emph{XMM} telescopes, the maps obtained were convolved with a
Lorentzian kernel of $\Gamma=7.125\arcsec$ corresponding to a
FWHM of the distribution of $6.6\arcsec$, the nominal FWHM of the
\emph{XMM} mirrors.

Finally, the convolved image was used to compute the encircled
PWN fraction in each extraction region of our \emph{XMM} image.

\subsection{Background Estimate From Spectral Fitting}
The \emph{XMM} spectra extracted from the four annuli
(Fig.~\ref{fig:spectra}) were fitted simultaneously using the following combination: \\


(interstellar absorption)$\times$(\,$\rho_i$(PSR model)\,+\,$\epsilon_i$(PWN model)\,) \\

where $\rho_i$ and $\epsilon_i$
are the PSR and PWN encircled fractions within the $i^{th}$ extraction region (see 
Sect.~\ref{par:psreef} and \ref{par:pwneef}). 
The interstellar absorption does not depend on $i$.
The PWN model is a power law whose index $\gamma_i$ is allowed to vary among
different regions, while  
the model describing the PSR emission does not depend on $i$. 
Our approach simultaneously resolve the PSR and PWN emission and 
compute best fit parameters for both the PWN and the PSR models (described in the following sections).

The spectral fitting was performed with XSPEC v.11.3, in the energy range $0.2-10\ \mathrm{keV}$.
The values of the coefficients are given in table~\ref{tab:coeff1}, and represent the different contribution of each region to the total flux of the point-like and diffuse emission. Uncertainties on the pulsar as well as on the nebular coefficients arise from the following caveats:
\begin{itemize}
\item{the pulsar spectrum is far from monochromatic;}
\item{the \emph{XMM-Newton} PSF is known within an uncertainty of $\sim2\%$ \citep{xmm_cal_status};}
\item{the PWN emission is known to be highly variable over a small time scale \citep{2001ApJ...554L.189P};}
\item{the Lorentzian kernel differs from the telescopes PSF;}
\item{the HRC efficiency differs from the pn one and also the telescopes effective
areas have different dependencies on the energy of the incoming
photons.}
\end{itemize}

Thus, to account for all uncertainties, an overall systematic
error of $5\%$ has been added.

\subsubsection{EPIC/pn Vela Pulsar Wind Nebula spectrum}
\label{par:pwn_spectrum}
The overall (absorption + PSR + PWN) 
best fit model, summarised in table~\ref{tab:fit}, allows us to determine 
the nebular background flux and photon index for all spectra. In particular we found 
that the inner circle contains  $7.60\pm0.04\ \mathrm{cts\ s^{-1}}$ from the PSR 
and $2.58\pm0.02\ \mathrm{cts\ s^{-1}}$ from the PWN 
(note that only $58\%$ of the pulsar counts are 
contained in this region - see table~\ref{tab:coeff1} and ~\ref{tab:coeff2}).

From table~\ref{tab:fit} we can see that the non-thermal nebular 
emission, whose photon index in the 4 concentric regions are
$\gamma_1$, $\gamma_2$, $\gamma_3$ and $\gamma_4$, become softer
as the distance from the pulsar increases. The result is
consistent with \emph{Chandra}-ACIS spectra obtained from 30
November 2000 observation \citep[see also ][]{2004IAUS..218..195K}.
The power-law normalisation yields the 
PWN flux within 40\arcsec from the PSR 
in the EPIC/pn
data, its value turns out to be  {\bf $F_\mathrm{X}=4.26 \pm 0.03\
\times 10^{-11} \mathrm{erg}\ cm^{-2}\ s^{-1}$ }, in the
range $1-8\ \mathrm{keV}$.
A large part of the PWN lies outside such region.
In order to estimate the flux of the remaining portion of the PWN inside 
the pn field of view, we extracted a spectrum from the whole detector, 
excluding the 40\arcsec circle
centered on the PSR, as well as a stripe affected by PSR out-of-time events.
Ad-hoc response matrix and effective area file were generated.
We fitted an absorbed power law model to the spectrum in the range 
$2-8\ \mathrm{keV}$, which yielded a best fit photon index of $1.61\pm0.02$
and an unabsorbed $1-8\ \mathrm{keV}$ flux of $(1.51\pm0.02)\times 10^{-11}\, \mathrm{erg}\
cm^{-2}\ s^{-1}$ ($\chi^2_{\nu}=0.99$, 140 d.o.f.). Thus,
the total flux of the PWN inside the pn field of view 
is $F_\mathrm{X}=(5.77 \pm 0.04)\ \times 10^{-11} \mathrm{erg}\ cm^{-2}\ s^{-1}$, 
corresponding to a luminosity of
\begin{equation}
L_\mathrm{X}=(5.74 \pm0.04)\ \times 10^{32} \mathrm{erg\ s^{-1}}
\end{equation}
at the parallactic distance ($\sim287\ \mathrm{pc}$). Such value is in good agreement with \cite{2002ASPC..271..181K}.

\subsection{EPIC/pn Vela pulsar spectrum}
\label{par:pa_spectrum}

The spectral distribution of the $\sim500\,000$ Vela PSR  photons
detected in the inner 10\arcsec{} radius circle, can be described
as the superposition of different components. A simple blackbody
thermal model does not fit well the observed spectra, that
appear harder than a simple planckian emission. The structure of
the residuals suggests the addition of a second ``pulsar''
component: both a second blackbody or a power-law gave acceptable
fits. Similar results were obtained with a \emph{magnetised
hydrogen atmosphere + power-law} model. The interstellar column
density of $\sim2.6\times 10^{20}\ \mathrm{cm^{-2}}$, inferred
from the different models, is in good agreement with the results
of \cite{2001ApJ...552L.129P} and \cite{2004AdSpR..33..503M}. For
the double blackbody model we obtained $\chi^2_{\nu}=1.1$ (1010
degrees of freedom). Assuming a parallactic distance of $287\
\mathrm{pc}$, the soft spectrum is described by a cooler
($T_\mathrm{bb}=1.06\pm0.03\times 10^6\ \mathrm{K}$) component,
with a radius $R_\mathrm{bb}=5.1^{+0.4}_{-0.3}\ \mathrm{km}$ and
a smaller and hotter one of
$T_\mathrm{BB}=2.16^{+0.06}_{-0.07}\times 10^{6}\ \mathrm{K}$ and
$R_\mathrm{BB}=0.73^{+0.09}_{-0.07}\ \mathrm{km}$. An equally
good fit was obtained with a $\gamma=3.48^{+0.08}_{-0.06}$
power-law; in this case the blackbody is found to have a
temperature $T=1.49^{+0.02}_{-0.02}\times 10^{6}\ \mathrm{K}$ and
a radius $R=2.0^{+0.6}_{-0.4}\ \mathrm{km}$. The power-law is
much steeper than that observed in all other X-ray emitting
pulsars.

A hydrogen atmosphere plus power-law model yields a  slightly
better fit ($\chi^2_{\nu}=1.0$ for 1010 degrees of freedom). The
surface temperature, for a radius $R=10\ \mathrm{km}$ and a
magnetic field $B=10^{12}\ \mathrm{G}$, is
$T_\mathrm{ha}=0.681\pm0.004\times 10^{6}\ \mathrm{K}$; the best
fit distance $D=269^{+12}_{-14}\ \mathrm{pc}$ agrees with the
parallattic measurements. The power-law photon index
$\gamma=2.8\pm0.2$ is however steeper that the one found by
\cite{2001ApJ...552L.129P}.

\begin{figure}
\includegraphics[angle=270,width=10cm]{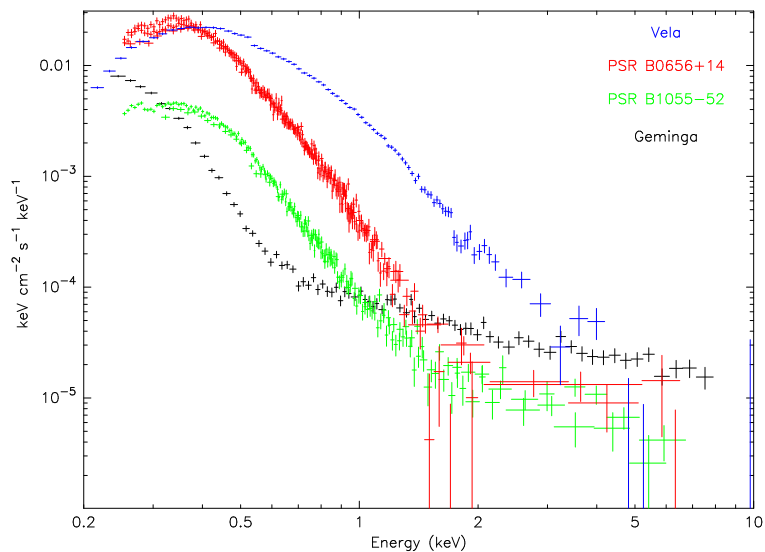}
\caption{\label{fig:3m_dart_eufs} \emph{XMM-Newton} spectra of the Three Musketeers and the \protect\object{Vela pulsar} \citep{2005ApJ...623.1051D}. The Vela pulsar thermal spectrum drops below the harder nebular emission at energies above $\sim$ 2 keV.}
\end{figure}

In Fig.~\ref{fig:3m_dart_eufs}, we plotted the Vela unfolded
spectrum  together with those of Geminga, PSR\ B1055-52 and PSR\
B0656+14 \citep{2005ApJ...623.1051D}. In the case of Vela we
subtracted the PWN contribution as estimated from spectral
fitting. Given the similarities between the four
\emph{XMM-Newton} pulsar spectra, we decided to analyse the Vela
data using  a phenomenological model encompassing two blackbody
and a power law component
\citep{2004Sci...305..376C,2005ApJ...623.1051D}.

\section{Timing Analysis}

The temporal resolution of the pn CCDs in ``Small Window''
readout mode is 5.6718 ms. Thus, the time-tagged XMM-pn photons
are perfectly suited to study the modulation of the Vela PSR flux
as a function of its $\sim$ 89 msec rotation period. The  photon
arrival times were corrected for the discrete sampling due to the CCD
readout and converted to the solar system
barycentre with the SAS task \emph{barycen}. Source counts were
extracted from the same region used for spectral analysis after
subtraction of the non point source contribution.

Epoch folding yields a strong signal at the pulsar frequency.
Following the prescription of \cite{1987A&A...180..275L} for high
accuracy period determination and error evaluation, we found
$P_{\mathrm{101}}=0.08933185\pm0.00000001\ \mathrm{s}$ and
$P_{201}=0.089331857\pm0.000000007\ \mathrm{s}$,
respectively for observation 0111080101 and 0111080201. Such
values agree with the contemporary radio ephemeris, kindly
provided by ATNF (R. Dodson, private communication) and reported
in table~\ref{tab:ef}. The temporal series were then merged and
folded using the radio $f$ and $\dot{f}$, to align in phase X-ray
and radio light curves.

\begin{figure}
\includegraphics[angle=270,width=10cm]{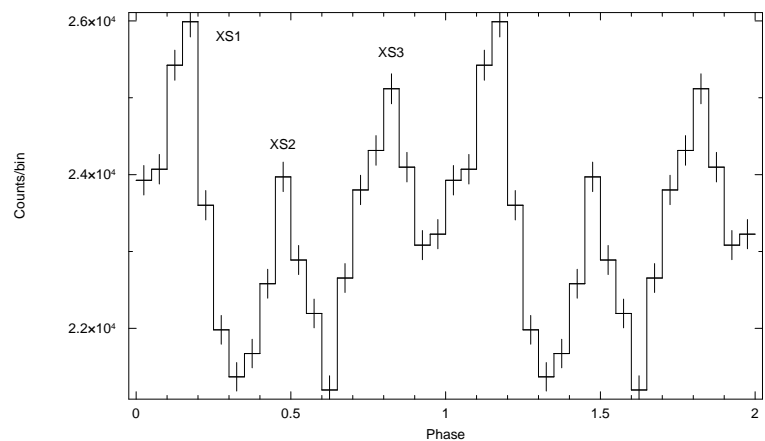}
\caption{\label{fig:curva_pn_tot} Folded light curve of the
\protect\object{Vela pulsar}  in the 0.2-10.0 keV
  energy range as observed with the EPIC/pn camera. Two rotational periods are
  represented for clarity. Phase 0 corresponds to the radio peak.}
\end{figure}

The overall (0.2--10 keV) X-ray light curve shows 3 broad peaks
per period marked $\mathrm{XS}_1$, $\mathrm{XS}_2$,
$\mathrm{XS}_3$ in Fig.~\ref{fig:curva_pn_tot}. The first and
highest peak is phase aligned with the first $\gamma$-ray peak
and follows the radio peak by $\sim0.15$ in phase. The second
one, the lowest in this energy range, reaches its maximum at
$\varphi=0.45-0.50$, and corresponds to the RXTE Peak 2-Soft
\citep{2002ApJ...576..376H} and UV $\mathrm{P}2_{\mathrm{s}}$
\citep{2005ApJ...627..383R}. The third peak, which has an
intermediate intensity, occurs at $\varphi=0.80-0.85$; it appears
in the soft X-ray light curve and is not present at any other
wavelength. The nebular background emission accounts for $25.5\%$
of the observed flux; the net count rate has a pulsed fraction of
$9.2\pm0.3\%$  (where pulsed fraction is defined as the ratio
between number of counts above the minimum and total number of
counts).



We note the presence of a pulsed signal also at energies above  2
keV, where the contribution from the two blackbody thermal
emission model used to describe Vela spectrum is negligible. This
would imply a non thermal origin for the pulsed emission above 2
keV, a component that was hidden by the nebular emission in the
phase-integrated spectrum. We estimated the 2--8 keV Vela light
curve to be about $100\%$ pulsed. However, we stress that, in such
an energy range, the source signal accounts for a small fraction
(less of $5\%$) of the photons detected, so that a small error in
the background estimate would translate into a big uncertainty in
the pulsed fraction.

\begin{figure}
\includegraphics[angle=0,width=10cm]{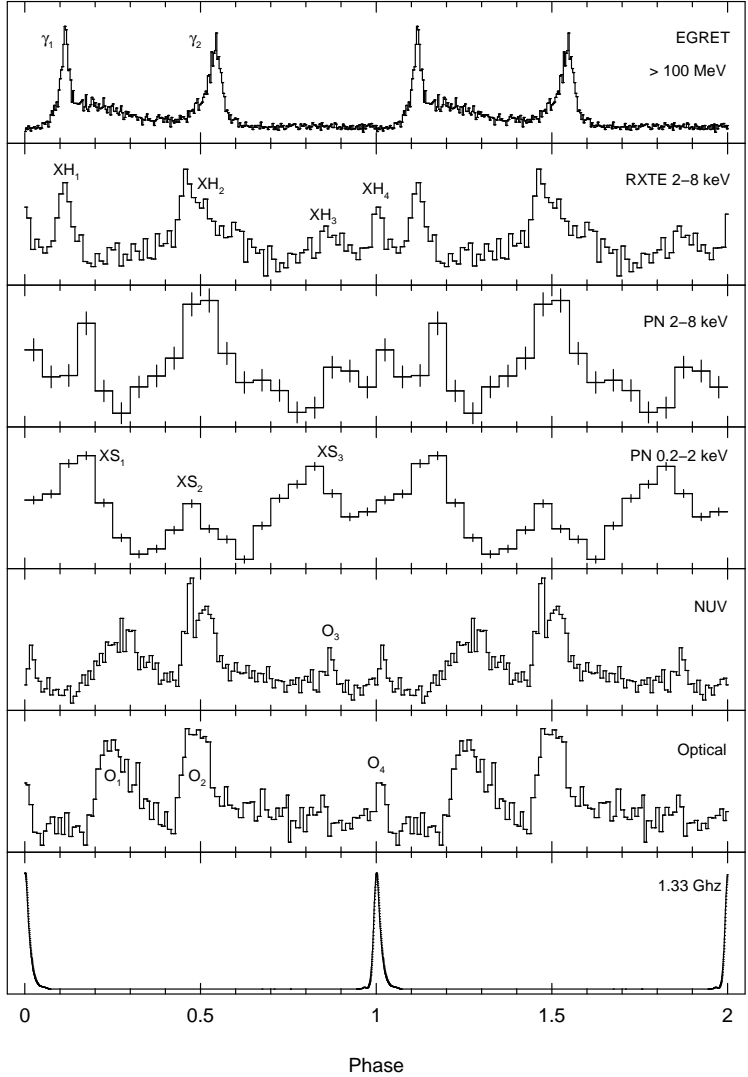}
\caption{\label{fig:curve_luce} Multiwavelength light curves of
the  \protect\object{Vela pulsar} from radio to $\gamma$-rays
\citep{2002ApJ...576..376H,2002nsps.conf...91K}. Two rotational
periods are represented for clarity. In the EPIC/pn light curves,
error bars are also plotted.}
\end{figure}

Fig.~\ref{fig:curve_luce} summarizes the multiwavelength behaviour
of the Vela pulsar. Significant differences are seen in the pulse
profile emerging from the two energy-resolved EPIC/pn light
curves: in the energy range 2--8 keV the second peak is stronger
than at lower energies and the third peak is not observable; the
phase interval $0.7<\varphi<1.2$ appears more complex at energies
above 2 keV, with 3 distinct peaks, two coinciding with the
optical third and fourth peak, and one phase aligned with the
first hard X-ray and $\gamma$-ray peak.

\section{Phase-resolved Spectral Analysis}

Since in the \emph{XMM-Newton} domain the \emph{pulse} shape  of
the \object{Vela pulsar} changes with energy, we expect a
variation of the \emph{spectral} shape as a function of the PSR
rotation. To study such a modulation we extract spectra from 20
different phase intervals. The spectra were rebinned in order to
have at least 30 counts per bin. Following the approach of
\cite{2004Sci...305..376C} and \cite{2005ApJ...623.1051D} we have
compared the phase-resolved spectra with the two blackbody best
fit model using the two normalisation coefficients as free
parameters.

\begin{figure}
\includegraphics[angle=270,width=10cm]{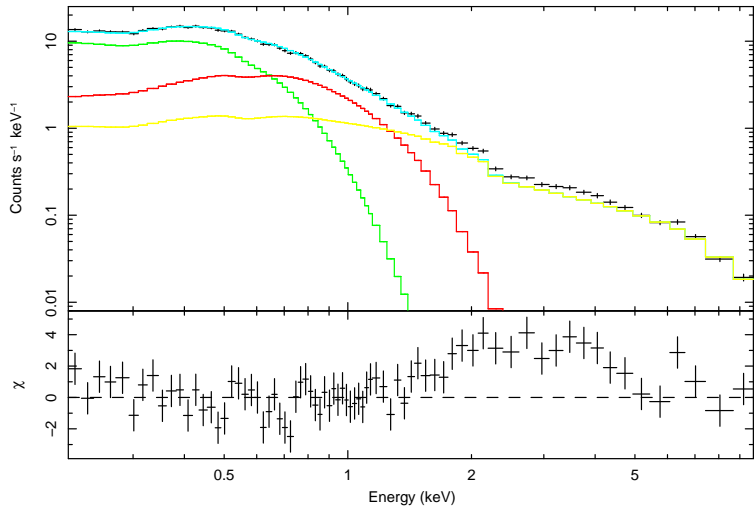}
\caption{\label{fig:phase_5_renorm} Observed
 spectrum for the
phase interval 0.45-0.55 and best fit double blackbody model 
folded with the instrument response. The two blackbody components of the pulsar
spectrum are shown (green curve: cool blackbody; red curve: hot blackbody)
together with the nebular contribution (yellow curve). The
lower panel shows the residuals in unit of standard deviation.
Above $\sim2$ keV, where the nebular contribution is
overwhelming, the data are systematically above the best fit
encompassing the double blackbody together with the nebular
contribution.}
\end{figure}

Such an approach works well for all the phase resolved  spectra
but for that encompassing phase interval $0.45<\varphi<0.55$ which
is characterized by high energy residuals, impossible to account
for with the two blackbody model. The spectrum of such a phase
interval is shown in Fig.~\ref{fig:phase_5_renorm}.

\begin{figure}[h!]
\includegraphics[angle=270,width=10cm]{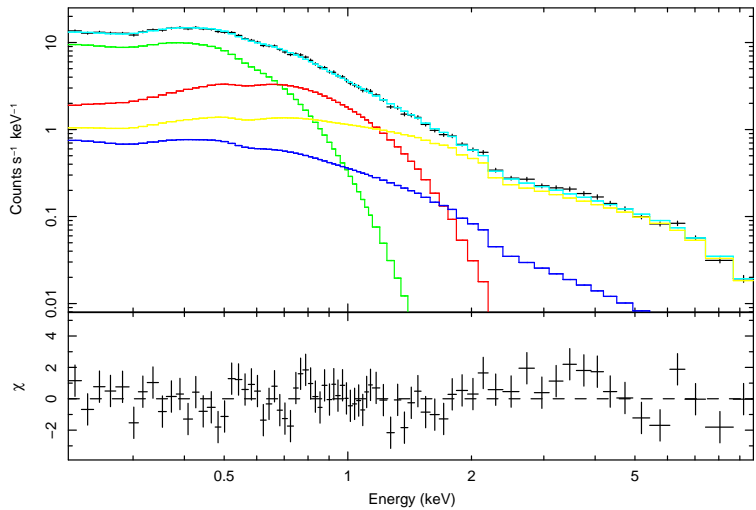}
\caption{\label{fig:phase_5_bf} As for
Fig.~\ref{fig:phase_5_renorm} but with a double blackbody plus
power-law model for the pulsar emission. Note the reduction of the high energy residuals.}
\end{figure}

Leaving the temperature of the two blackbodies and the hydrogen
column density fixed to the values found above, a power-law with
photon index $2.2 ^{+0.4} _{-0.3}$ is required to fit the spectrum
(Fig.~\ref{fig:phase_5_bf}) for the $0.45<\varphi<0.55$ interval. The normalization of the power-law
is $\sim10$ times lower than the nebular emission in this phase
interval, thus, its contribution to the phase-integrated spectrum
is $\sim100$ times lower than the nebular one and it cannot be
seen in the total Vela spectrum.

Following the detection of a non-thermal component, we decided  to perform spectral fitting with a three
component  source model (two blackbodies + power-law) plus the
background model to account for the nebular emission. As in the
previous case, the blackbody temperatures where fixed to the
values best fitting the total Vela spectrum
 while a power-law
photon index of 2.2 was used. The best fitting parameters are
given in table~\ref{table:phase_res_norm} while an animated version of the
phase resolved spectra 
can be found in http://www.iasf-milano.inaf.it/$\sim$deluca/vela/.

\begin{figure}
\includegraphics[angle=270,width=13cm]{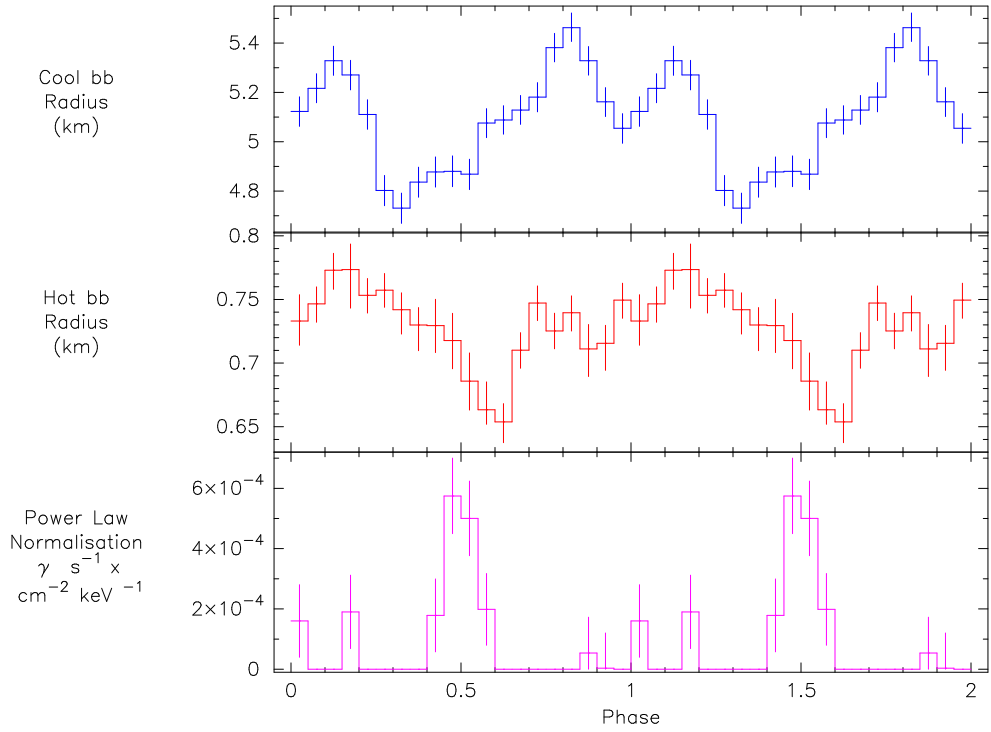}
\caption{\label{fig:phase_res} Variation of the three spectral
component  (cool and hot blackbody radius and power-law
normalization) as a function of the rotational phase. Two
rotational periods are plotted for clarity.}
\end{figure}

From the blackbody emissions, we can compute blackbody radii (at
parallactic distance). In Fig.~\ref{fig:phase_res} we reported
the variation of blackbody radii and power-law normalization as a
function of the rotational phase. The hotter blackbody presents a
low modulation, with a single broad peak per period. The cooler
one shows a more complex modulation with possibly two peaks per
period, with a small ``dip'' between them, well aligned with the
radio pulse. The cooler component first maximum (R$\sim5.3\
\mathrm{km}$) trails the radio pulse phase by about $10\
\mathrm{ms}$ ($\sim0.10-0.15$ in phase). After such maximum, the emitting
area suddenly decreases to reach its minimum (R$\sim4.7\ \mathrm{km}$),
at $\varphi\simeq0.30$, the same phase of the absolute minimum of
the Vela soft X-ray flux; the cool blackbody radius then slowly
grows and reaches its second and more pronounced maximum
(R$\sim5.5\ \mathrm{km}$) at phase 0.8-0.85 responsible for the
third \emph{XMM} light curve peak which is not seen in hard X-rays, nor in the ultraviolet or optical light
curves.

The hotter blackbody peaks, with a maximum radius of about $0.77\
\mathrm{km}$, at the same phase of the first cool blackbody peak, trailing by
0.1-0.15 in phase the radio pulse, long known to mark the pulsar polar region. The minimum of the hot blackbody emission occurs at about 180\degr\ from the maximum and trails the minimum of the cooler emission by $\sim0.3$ in phase. The transition from the minimum to the maximum state is not symmetric, as the growth is sharper than the descent.


The non-thermal emission is present with a narrow peak between $\varphi\sim0.4$ and $\varphi\sim0.6$, between the two thermal emissions minima. The power-law normalisation is found to be consistent with zero in all other phase intervals. Such non thermal component is responsible for the second peak observed in the 2--8 keV EPIC light curve and seems to be connected with the narrow ``spike'' in the NUV/FUV light curves and with the ``peak 2-Soft'' detected with \emph{RXTE}. The power-law photon index also agrees with the value found in \emph{RXTE} phase-resolved spectroscopy \citep{2002ApJ...576..376H} in the same phase interval. The peak corresponds to the ``Leading Wing 2'' in the EGRET energy range.


\section{Discussion}
Considering the different spectral components identified in the EPIC/pn data we note that the maximum extension of the hotter thermally emitting region is observed when looking at the polar region of the neutron star. The radius of this blackbody component is in agreement with the radius of a polar cap within a simple magnetic dipole model \citep{1969ApJ...157..869G}. For the rotational velocity of the \object{Vela pulsar}, the polar cap radius would be $R_{\mathrm{PC}}=R(R\Omega/c)^{1/2}=0.485\ \mathrm{km}$. \cite{2002ApJ...568..862H} made an estimate of the energetics for an \emph{hot spot} re-heated by curvature particle downflow: with the Vela parameters the bolometric luminosity would be $L=6.01\times 10^{31}\ \mathrm{erg\ s^{-1}}$ and the temperature $T=2.66\times 10^{6}\ \mathrm{K}$. Both values agree with what we found. Since the \object{Vela pulsar} is known to be an inclined rotator, a terrestrial observer would face a single polar region during rotation. Assuming that the hot component is produced by the polar cap region heated by return current, the gravitational bending \citep{1995ApJ...442..273P}
would result in a shallow modulation of the emitting surface, in agreement with our results.

The cooler blackbody would represent the radiation from the remaining part of the star surface.
The radius inferred from the phase-averaged spectrum is too small to fit in any proposed equation of state for a star composed mainly of neutron. We also observe a $\sim$10$\%$ modulation of this spectral component as a function of the rotational phase. Magnetospheric reprocessing of the thermal photons emitted from the surface could provide a phase-dependent ``obscuration'' of a fraction of the neutron star surface, depending on magnetic field configuration and viewing geometry. The phenomenon of the magnetospheric ``blanket'', e.g. cyclotron resonance scattering by plasma at a few stellar radii \citep{2003astro.ph.10777R}, originally proposed by \cite{1993ApJ...415..286H} as an explanation of the soft thermal emission of Geminga, could provide the physical basis for the observation of a phase-dependent emitting area.

Anisotropic heat transfer from the interior of the neutron star, would also
 provide a surface temperature far from uniform, with the polar regions hotter
 than the equatorial ones \citep{1983ApJ...271..283G}. Such non-uniform
 temperature distribution would result into a modulated X-ray thermal
 flux. Alternatively, using a uniform surface temperature approximation, this
 translates into a modulation of the emitting
 surfaces. \cite{1995ApJ...442..273P} show that in such a case a modulation of
 a few percent is expected, owing to gravitational light bending. However such
 a modulation would have to be phase aligned with that of the hotter component.
The complicate modulation we observe for the cool blackbody component is not easy
to reconcile with such a picture, pointing to a more complex surface
 temperature distribution, and/or to magnetospheric reprocessing as discussed above.
%
%

The non-thermal component detected in a small phase interval 180\degr\ from the radio pulse, when the thermal components are at their minimum, might be produced by particles accelerated in the star magnetosphere.

\section{Conclusions}


The spectral analysis of the \emph{XMM-Newton} data of the Vela
pulsar  confirms the thermal nature of its emission pointing
towards an X-ray phenomenology similar to that seen in older
neutron stars. Both the overall spectral shape and the
phase-resolved behaviour are reminiscent of what has been found
for middle aged pulsars such as Geminga, \object{PSR\ B0656+14}
and \object{PSR\ B1055-52}, affectionately called the ``Three
Musketeers'' on view of their similarities \citep{1997A&A...326..682B}. Indeed, the Three
Musketeers could be described, even with some caveats, with a
simple common phenomenological model in which the differences in
the variations of the ``hot spot'' emitting areas as a function
of the rotational phase could be ascribed to the different
viewing geometries of the three neutron stars which are known to
be different: orthogonal rotator seen perpendicular to the
rotational axis for \object{PSR\ B1055-52} and almost aligned for
\object{PSR\ B0656+14} \citep[see e.g.][for a full description of
such a picture.]{2005ApJ...623.1051D}.

While Vela is certainly hotter than the older musketeers, its
overall blackbody emitting radius is rather small leading to a
low luminosity, well below the value expected in the standard
cooling scenario described by \cite{2002ApJ...571L.143T}. Using a
hydrogen atmosphere eases the radius problem but, yielding lower
temperature, does not change the source luminosity which rests on
a precisely determined parallactic distance.

Moreover, although Vela is known to be an inclined rotator \citep[the
angle between the magnetic dipole vector and the rotation  axis
is $\alpha\sim60\degr$, while the line of sight is inclined with
respect to the latter by  $\zeta\sim55\degr$,][ and references 
therein]{2001ApJ...556..380H}, our phase
resolved spectroscopy points to a surprisingly small modulation,
typical of aligned rotators, such as \object{PSR\ B0656+14},
rather than orthogonal ones.

The presence of a non-thermal component sharply pulsed in
antiphase with the radio pulse is a surprising result of our
analysis and could be exploited to better constrain the geometry
of the emitting regions with the goal to unveil the emission
mechanism behind Vela complex multiwavelength behaviour.

The remarkable body of evidence collected so far on the  Vela
pulsar has been generally interpreted in the framework of the
outer gap model \citep{1995ApJ...438..314R} where e$^{+/-}$
couples fill the pulsar magnetosphere high above the star
surface. Indeed, such a model predicts an X-ray non-thermal broad
peak at $\varphi\sim0.35-0.5$, a feature clearly seen both in
\emph{RXTE} and \emph{XMM} light curves.

Recently, the Vela pulsar phenomenology has been accounted  for
also by the two pole caustic model \citep{2003ApJ...598.1201D} to
supply the high energy particles.

\emph{XMM-Newton} findings provide a useful link between the star thermally emitting surface and its particle filled magnetosphere. Such two-regime, phase-resolved view probes the star geometry as it rotates and provides inputs for future tri-dimensional realistic models of the pulsar magnetosphere.

\acknowledgments

We thank R.Dodson for providing Vela pulsar radio ephemeris  
simultaneous with the {\emph XMM-Newton} observations.
We thank A.Possenti for his help with the radio data.
We would also thank A.Harding and M.Strickman for their help
to understand the RXTE phase alignment.
XMM-Newton and Chandra data analysis is supported 
by the Italian Space Agency (ASI) under contract ASI/INAF I/023/05/0. 
ADL acknowledges an ASI fellowship.



\clearpage

\begin{table*}
\caption{Journal of \emph{XMM-Newton} observations.}
\label{table:time}
\centering
\begin{footnotesize}
\begin{tabular}{ccccccc}
\tableline \tableline
 Obs ID & Date & MJD & Obs. Time (ks) & Instrument(mode) & Good Time (s) & Live Time (s)\\
\hline
0111080101 & 2000 Dec 1 & 51879 & 41.11 & pn(SW) & 37\,590 & 26\,366 \\
           &           &       &        & MOS1(SW)& 37\,648 & 37\,241\\
           &           &       &        & MOS2(LW)& 37\,694 & 36\,597 \\
0111080201 & 2000 Dec 2 & 51880 & 61.81 & pn(SW) & 50\,887 & 35\,693 \\
           &           &       &        & MOS1(SW)& 49\,067 & 48\,524 \\
       &           &       &        & MOS2(LW)& 50\,410 & 48\,940 \\
\tableline
\end{tabular}
\end{footnotesize}
\tablecomments{For the two different XMM datasets are listed the start date of the observation in standard UTC time and in Modified Julian Day (MJD), the total observation time, the instrument that performed the observation and the operational mode (SW=``Small Window'', LW=``Large Window''), the good time intervals after removing the Soft Proton Flares, the Live Time after correction for charge transferring and readout time.}
\end{table*}

\begin{table}
\caption{2D best fit counts in the annular region 2.64\arcsec--3.43\arcsec for the three HRC-I images.}
\label{tab:PWN}
\begin{center}
\begin{tabular}{lccc}
\hline
\hline
Obs. Id      &           364          &      1518          &      1966        \\
\hline
$\mathrm{cts\ pixel^{-2}}$ &    $1.43 \pm0.04$  & $1.41\pm0.04$ & $1.23\pm0.04$ \\
\hline
$\mathrm{cts\ pixel^{-2}\ s^{-1}}$&    $0.0017$            & $0.0016$           & $0.0014$          \\
\hline
\end{tabular}
\end{center}
\end{table}

\begin{table}
\caption{Total number of counts.}
\label{tab:coeff1}
\begin{center}
\begin{tabular}{c c c c}
\hline\hline
\rule[-5pt]{0pt}{4ex}
Extraction  & Total number & PSR EEF & PWN flux \\
region & of photons & ($\%$) & (counts) \\
\hline
\rule{0pt}{3ex}
       10\arcsec            & 671\,021 & 0.582342  & 17729.6985 \\
    10\arcsec--20\arcsec    & 541\,089 & 0.215818  & 40962.9075 \\
    20\arcsec--30\arcsec    & 431\,813 & 0.0770648 & 42880.3005 \\
    30\arcsec--40\arcsec    & 293\,991 & 0.0377001 & 29622.9375 \\
\hline
\end{tabular}
\tablecomments{Total number of photons collected by the EPIC/pn in the different spectra extraction regions. Also shown are \emph{XMM-Newton} EEF and 
the PWN counts estimated on the Chandra/HRC surface brightness map (degraded to the XMM angular resolution). The PWN parameters
$\epsilon _i$ used in the fit are proportional to such counts. We assumed $\epsilon _1$=1. 
These values have been used to compute the PSR and PWN coefficients for the spectral fitting.}
\end{center}
\end{table}

\begin{table}
\caption{Pulsar and Nebula contributions.}
\label{tab:coeff2}
\begin{center}
\begin{tabular}{c c c c}
\hline\hline
\rule[-5pt]{0pt}{4ex}
Extraction  &  PSR flux & PWN flux & EPIC/pn counts\\
region &  (counts) & (counts) & (0.2-10 keV)\\
\hline
\rule{0pt}{3ex}
        10\arcsec            & 471\,592      &      160\,114       & 631711\\   
    10\arcsec--20\arcsec    & 164\,272      &       355\,540     &519824\\   
    20\arcsec--30\arcsec    & 54\,544      &      363\,174      &417689 \\   
    30\arcsec--40\arcsec    & 36\,491      &      247\,743      & 284238\\   
\hline
\end{tabular}
\tablecomments{Number of photons collected by the EPIC/pn in the different spectra extraction regions, emitted from the Vela pulsar and from its PWN, as derived from best fit spectral model fluxes.}
\end{center}
\end{table}

\begin{table}
\caption{Best fit parameter for double blackbody model.}
\label{tab:fit}
\begin{center}
\begin{tabular}{p{70pt}c}
\hline\hline
{\bf{Parameter}} & \multicolumn{1}{c}{\bf{Best fit value}}\\
\hline
\multicolumn{2}{c}{\rule{0pt}{3ex} {\small Cool blackbody (bb) + Hot blackbody (BB)}}\\
\hline
$nH~(\times 10^{22}~cm^{-2})$  & $(2.59\pm0.01)\times 10^{-2}$ \\  
$k_{B}T_\mathrm{bb}$ (keV)\dotfill  & $0.091\pm0.003$ \\  
$norm_\mathrm{bb}$  \dotfill  &$18000^{+3000} _{-2000} $  \\ 
$k_{B}T_\mathrm{BB}$ (keV) \dotfill     &$0.186^{+0.005} _{-0.006}   $ \\ 
$norm_\mathrm{BB}$ \dotfill   &$380 ^{+90} _{-70} $ \\ 
$\gamma_1$  \dotfill   &$1.367 ^{+0.006} _{-0.007}      $ \\ 
$\gamma_2$ \dotfill    &$1.495 ^{+0.006} _{-0.006}        $ \\ 
$\gamma_3$ \dotfill    &$1.599 ^{+0.006} _{-0.006}  $ \\ 
$\gamma_4$  \dotfill   &$1.651 ^{+0.007} _{-0.006} $ \\ 
$norm_\mathrm{PWN}$ \dotfill  & $(1.033\pm0.006)\times 10^{-3}$ \\ 
\hline
\end{tabular}
\tablecomments{Best fit parameter for the PSR double blackbody model and for
  the PWN. Spectra from the four regions described in Sect.~\ref{par:spectra} have been used.
$nH$ is computed with a photo-electric absorption model (model {\em wabs} in
XSPEC) using Wisconsin \citep{1983ApJ...270..119M} cross-sections and the 
relative abundances by \citet{1982GeCoA..46.2363A}.
Blackbody normalizations are defined as (R/D$_{10}$)$^2$, where R is the 
emitting radius in km and D$_{10}$ is the distance in units of 10 kpc.
The PWN photon indices $\gamma_1$, $\gamma_2$, $\gamma_3$ and $\gamma_4$ are the best fit values
for the four extraction regions, from the innermost circle to the outermost annulus (see text).
As explained in the text, 
a common PWN normalization parameter is simultaneously fitted to all spectra,
since the PWN flux fraction in different region is computed {\em a priori}
using the Chandra surface brightness map (degraded at the XMM angular resolution) and is kept fixed during the fit.
The normalization of the PWN in the $i^{th}$ region is obtained by multiplying 
the value quoted in the table (in units of
photons cm$^{-2}$ s$^{-1}$ keV$^{-1}$ at 1 keV) by the PWN parameter $\epsilon_i$ (see text and table~\ref{tab:coeff1}).
Errors are computed at a $90\%$ confidence level for a single interesting parameter.}
\end{center}
\end{table}

\begin{table}
\caption{Physical characteristics of the vela pulsar}
\label{table:bf_fisico}
\begin{center}
\begin{tabular}{p{60pt}c}
\hline\hline
{\bf{Parameter}} & \multicolumn{1}{c}{{\bf Best fit value}}\\
\hline
\multicolumn{2}{c}{\rule{0pt}{3ex} {\small Cool blackbody (bb) + Hot blackbody (BB)}}\\
\hline
$T_\mathrm{bb}\ (\mathrm{K})$      \dotfill&$(1.06\pm0.03)\times 10^{6}  $ \\ 
$R_\mathrm{bb}\ (\mathrm{km})$  \dotfill&$5.06 ^{+0.42} _{-0.28}  $ \\ 
$L_\mathrm{bb}\ (\mathrm{erg\ s^{-1}})$ \dotfill&$(2.3 ^{+0.5} _{-0.4})\times 10^{32} $ \\ 
$T_\mathrm{BB}\ (\mathrm{K}) $\dotfill&$(2.16 ^{+0.06} _{-0.07})\times 10^{6}  $ \\ 
$R_\mathrm{BB}\ (\mathrm{km})$   \dotfill&$0.73 ^{+0.09} _{-0.07}    $ \\ 
$L_\mathrm{BB}\ (\mathrm{erg\ s^{-1}})$ \dotfill&$(8.3^{+2.2} _{-1.8})\times 10^{31} $ \\ 
$L_\mathrm{B}^\mathrm{tot}\ (\mathrm{erg\ s^{-1}})$ \dotfill&$(3.1 ^{+0.5} _{-0.4}) \times 10^{32} $ \\ 
$L^\mathrm{PWN}_\mathrm{X}\ (\mathrm{erg\ s^{-1}})\footnotemark[1]$ \dotfill& $(5.74 \pm 0.04)\times 10^{32} $ \\ 
\hline
\end{tabular}
\tablecomments{Physical characteristics of the \object{Vela pulsar} as inferred from the cool blackbody (bb) plus hot blackbody (BB) best fit model (see table \ref{tab:fit}). We assumed $58\%$ of the energy emitted encircled in the 10\arcsec{} radius region. 
Emitting radii are computed for a distance of 287 pc. 
The bolometric blackbody luminosities ($L_{\mathrm{bb}}$, $L_{\mathrm{BB}}$
and $L^{\mathrm{tot}}_{\mathrm{B}}$) are computed from Stefan-Boltzmann law,
for best fit temperature and radius. The PWN luminosity
($L^{\mathrm{PWN}}_{\mathrm{X}}$) is computed for the energy range 1--8 keV at
the parallactic distance, using the flux measured within the entire pn field of
view (see Sect.\ref{par:pwn_spectrum}).}
\end{center}
\end{table}

\begin{table}
\caption{Vela radio ephemeris (R. Dodson, private
communication).} \label{tab:ef}
\begin{center}
\begin{tabular}{p{80pt}l}
\hline \hline
$t_{0}$ in MJD \dotfill      & $51881.000000049$  \rule{0pt}{3ex}\\
$f\ (\mathrm{Hz})$         \dotfill  & $11.1942146219182$ \\
$P\ (\mathrm{s})$          \dotfill  & $0.0893318588016$  \\
$\dot f\ (\mathrm{Hz\,s^{-1}})$\dotfill& $-1.56297\times 10^{-11}$  \\
$\dot P\ (\mathrm{s \ s^{-1}})$\dotfill& $1.24728\times 10^{-13}$  \\
\hline
\end{tabular}
\end{center}
\end{table}

\begin{table}
\caption{Results of phase resolved spectroscopy}
\label{table:phase_res_norm}
\begin{center}
\begin{tabular}{ccccc}
\hline\hline
\rule{0pt}{3ex} Phase & bb radius & BB radius & PL norm & $\chi^2 _{\nu}$
(d.o.f.) \\ & (km)&(km)&($\gamma$ s$^{-1}$ cm$^{-2}$ keV$^{-1}$)\\
\hline
0.00-0.05  &  $5.12\ \pm\ 0.06$  &  $0.73 \pm 0.02 $  &  $(1.6 \pm
1.2)\times10^{-4}$ & 1.00 (136) \\
0.05-0.10  &  $5.22\ \pm\ 0.06$  &  $0.75  \pm 0.01 $  &  0 & 0.99 (137) \\
0.10-0.15  &  $5.33\ \pm\ 0.06$  &  $0.77  ^{+0.01} _{-0.02} $  &  0 & 1.10
(137)   \\
0.15-0.20  &  $5.27\ \pm\ 0.06$  &  $0.77  ^{+0.02} _{-0.03} $  &  $(1.9 \pm
1.2) \times10^{-4}$  & 1.10 (143)  \\
0.20-0.25  &  $5.11\ \pm\ 0.06$  &  $0.75  \pm 0.01 $  &  0 & 1.28 (137) \\
0.25-0.30  &  $4.80\ \pm\ 0.06$  &  $0.76  \pm 0.01$  &  0 & 1.41 (133) \\
0.30-0.35  &  $4.73\ \pm\ 0.06$  &  $0.74   ^{+0.01} _{-0.02} $ &  0  & 1.10
(134) \\
0.35-0.40  &  $4.84\ \pm\ 0.06$  &  $0.73   ^{+0.01} _{-0.02} $ &  0 & 1.51
(138)  \\
0.40-0.45  &  $4.88\ \pm\ 0.06$  &  $0.73  \pm 0.02$  & $(1.8 \pm
1.2)\times10^{-4}$ & 1.17 (138)   \\
0.45-0.50  &  $4.88\ \pm\ 0.06$  &  $0.72  \pm 0.02 $  & $(5.7 \pm
1.2)\times10^{-4}$ & 0.84 (140)  \\
0.50-0.55  &  $4.87\ \pm\ 0.06$  &  $0.69  \pm 0.02$ &
$(5.0\pm1.2)\times10^{-4}$ & 1.13 (144)  \\
0.55-0.60  &  $5.08\ \pm\ 0.06$  &  $0.66  ^{+0.02} _{-0.01} $  &  $(2.0 \pm
1.2)\times10^{-4}$ & 1.03 (139) \\
0.60-0.65  &  $5.09\ \pm\ 0.06$  &  $0.65   ^{+0.01} _{-0.02} $  &  0 & 1.08 (138) \\
0.65-0.70  &  $5.13\ \pm\ 0.06$  &  $0.71  \pm 0.01 $  &  0 & 1.08 (137) \\
0.70-0.75  &  $5.18\ \pm\ 0.06$  &  $0.75  \pm 0.01 $  &  0 & 1.26 (136) \\
0.75-0.80  &  $5.38\ \pm\ 0.06$  &  $0.73  \pm 0.01 $  &  0 & 1.08 (134) \\
0.80-0.85  &  $5.46\ \pm\ 0.06$  &  $0.74  \pm 0.01 $  &  0 & 1.09 (133) \\
0.85-0.90  &  $5.33\ \pm\ 0.06$  &  $0.71  \pm 0.02 $  &  $(5.0 _{-5}
^{+10})\times10^{-5}$ & 1.27 (134) \\
0.90-0.95  &  $5.16\ \pm\ 0.06$  &  $0.72  ^{+0.01} _{-0.02} $  & $ (0.3
_{-0.3} ^{+10}) \times10^{-5}$ & 1.14 (136)   \\
0.95-1.00  &  $5.05\ \pm\ 0.06$  &  $0.75  \pm 0.01 $  &  0 & 1.11 (134) \\
\hline
\end{tabular}
\tablecomments{Cool blackbody (bb), hot blackbody (BB) radius and power-law 
normalisation as function of the rotational phase. 
Temperatures of both blackbody components were fixed at the best fit 
values for the phase-integrated spectrum. The power law photon index was left as a 
free parameter in the 0.45-0.55 phase interval (best fit value of $2.2 ^{+0.4} _{-0.3}$).
For the remaining part of the cycle the photon index was fixed at the best 
fit value for the 0.45-0.55 phase interval.  For the radius computation we 
assumed $58\%$ of the energy emitted encircled in the 10\arcsec radius region 
and a parallattic distance of 287 pc. Errors are computed at a $90\%$ 
confidence level for a single interesting parameter.}
\end{center}
\end{table}

\clearpage

\end{document}